\documentclass[10pt,letterpaper,onecolumn]{article}
\usepackage{amsmath,amssymb,graphicx,cite,url,float,dsfont}
\usepackage{lettrine}
\usepackage{lineno}
\begin{document}
%\linenumbers
\title{\textsf{\textbf{Super sub-Nyquist single-pixel imaging by means of cake-cutting Hadamard basis sort}}}% Declares the document's title.

\author{Wen-Kai Yu$^1$}
\footnotetext[1]{Center for Quantum Technology Research, School of Physics, Beijing Institute of Technology, Beijing 100081, China. Correspondence and requests for materials should be addressed to W.-K.Y. (email: yuwenkai@bit.edu.cn)}

%\author{\textsf{\textbf{Wen-Kai Yu}}\renewcommand{\thefootnote}{\arabic{footnote}}\footnotemark[1] \renewcommand{\thefootnote}{\fnsymbol{footnote}}\footnotemark[1]}      % Declares the author's name.
%\renewcommand{\thefootnote}{\arabic{footnote}}
%\footnotetext[1]{Center for Quantum Technology Research, School of Physics, Beijing Institute of Technology, Beijing 100081, China}
%\renewcommand{\thefootnote}{\fnsymbol{footnote}}
%\footnotetext[1]{yuwenkai@bit.edu.cn}
\date{}% Deleting this command produces today's date.

\maketitle% Produces the title.

\renewenvironment{abstract}{%
    \setlength{\parindent}{0in}%
    \setlength{\parskip}{0in}%
    \bfseries%
    }{\par\vspace{-6pt}}

\begin{abstract}
Single-pixel imaging via compressed sensing can reconstruct high-quality images from a few linear random measurements of an object/scene known \textit{a priori} to be sparse or compressive, by using a point/bucket detector without spatial resolution. Nevertheless, it still faces a harsh trade-off among the acquisition time, the spatial resolution and the signal-to-noise ratio. Here we present a new compressive imaging approach with use of a strategy called cake-cutting which optimally reorders the deterministic Hadamard basis. By this means, the number of measurements can be dramatically reduced by more than two orders of magnitude. Furthermore, by exploiting the structured characteristic of the Hadamard matrix, we can accelerate the computational process and simultaneously reduce the memory consumption of storing the matrix. The proposed method is capable of recovering an image of the object, of pixel size $1024\times1024$, with a sampling ratio of even 0.2\%, thereby realizing super sub-Nyquist sampling and significantly reducing the acquisition time. Moreover, through the differential modulation/measurements, we demonstrate this method with a single-photon single-pixel camera under low light condition and retrieve clear images through partially obscuring scenes. This described practical method complements the single-pixel imaging approaches and can be applied to a variety of fields, such as video, night vision goggles and automatic drive.
\end{abstract}

\section*{\textsf{Introduction}}
\lettrine[lines=2]{T}{he} ability to capture the two-dimensional (2D) image information is extremely significant in plenty of applications, ranging from astronomical observation \cite{Kastner2002}, phase retrieval \cite{YuOC2017} to hyperspectral imaging \cite{Studer2012}. However, the sampling specified by the Nyquist-Shannon criterion for all pixels of an object image generally involves the acquisition of a huge amount of information, and is also accompanied by the questions of both transmission and storage. In addition, imaging with high spatial resolution needs advisable compromises on the conflicting goals of acquiring high signal-to-noise (SNR) for each pixel-unit whilst keeping low acquisition time. For example, in order to obtain a higher SNR, one can either increase the integration time \cite{Woringer2017}, just as in pixelated array/scanning detectors, thereby increasing the acquisition time, or gather the total luminous flux together, just as in single-pixel systems, but sacrificing the sampling time and the computation time.

To our knowledge, single-pixel imaging (SPI) might trace back to early raster scanning schemes, like the flying-spot camera in 1884, optical coherence tomography \cite{Huang1991} in 1991. As an alternative, it is also possible to calculate the intensity correlation between the random-modulated illumination patterns and the detected bucket signals, based on a statistical mechanism called ghost imaging (GI) \cite{ShihPRL1995,Boyd2002,Shapiro2008,Zhao2012,YuOE2014}. But in traditional GI, in order to obtain a nice reconstruction, the number of measurements should be considerably higher than the pixel dimension $N$ of the object image. Hadamard \cite{Duran2012,Clemente2013,MJSunNC2016,Huynh2016,Lochocki2016} and Fourier \cite{Zhang2015,Bian2016,Jiang2017} SPI are other two techniques that use complete deterministic orthogonal bases, allowing one to reduce the number of measurements to $N$. Since there is no need to use any pixelated array, SPI schemes can largely improve the flux, which is a benefit especially under ultra-weak light conditions. Additionally, they can work with low-cost at non-visible wavelengths, where array cameras are expensive and not well developed \cite{Radwell2014}.

Recently, some SPI methods based on compressed sensing (CS) \cite{Donoho2006,Candes2006,Candes2008,Baraniuk2008} have been proposed to acquire a better performance by exploiting the sparsity of the object. It enables a considerable decrease in the number of measurements without compromising the SNR, but at the cost of increased computational time (in minutes or hours). We also note that, for the general cases, one cannot acquire a good image quality when the sampling ratio is below 30\%. Many CS applications in fields including magnetic resonance imaging \cite{Lustig2008}, astronomy \cite{Bobin2008}, and microscopic imaging \cite{Shin2017,YuOC2016}, have resolutions generally smaller than $128\times128$ pixels. Besides, the larger the pixel resolution, the more stringent the computational restrictions, which is unsuitable for practical real-time imaging.

In terms of imaging mechanisms with few measurements, differential GI \cite{Ferri2010}, which is an early differential SPI scheme, enhances the SNR by subtracting the background noise. Then correspondence imaging (CI) has been proposed by Luo et al. \cite{Luo2011,Luo2012} and explained from many perspectives \cite{Shih2011,YuCPB2015,Yao2015}, allowing one to obtain a positive or negative image by conditional averaging the reference patterns, but without correlation calculation. As a result, the number of measurements is cut by at least half. Although diverse CI-based schemes \cite{LiAPL2013,MJSunAO2015,GLLi2016,Wu2017} have been developed, the performance is not greatly improved. Then, it is found that based on the differential measurements of random binary patterns and their inverse \cite{BSunSci2013}, a ghost image can be reconstructed by correlating non-inverted patterns. Note that the inverse patterns also can be utilized, a technique called complementary compressive imaging \cite{YuSR2014} and its derivative named differential CS \cite{YuOC2016} have been proven to produce a satisfactory image quality in orders of magnitude better than conventional CS or GI, with a sampling ratio around 15\%. More recently, an approach based on a ``Russian Dolls" (RD) ordering approach \cite{MJSunSR2017} is proposed to yield a comparable quality compared to CS at 6\% sampling ratio but the spatial resolution is still limited.

Here we present a single-pixel compressive imaging technique that can acquire high-quality images of large pixel scale $1024\times1024$ with super sub-Nyquist sampling ratio even below 0.2\%, by using a strategy that we call the ``cake-cutting" (CC) sort to optimally reorder the deterministic Hadamard basis. From the aspect of its physics nature, the sorting is based on the contribution of the pattern basis to the reconstruction, which is inspired from CI. Following this idea, the most significant patterns are always modulated firstly. Meanwhile, by utilizing the structured characteristic of Hadamard matrix, the computational overhead and the memory consumption is greatly reduced. It is shown through numerical simulation to significantly reduce the number of measurements along with the acquisition time. Furthermore, with a single-photon single-pixel camera setup based on differential modulation, we demonstrate its ability to retrieve clear images through partially obscuring scenes under noisy environmental illumination conditions.

\section*{\textsf{Results}}
\textbf{Principles description.} The core idea of single-pixel imaging is to shift the spatial resolution away from the array detectors and onto the modulated patterns, which are typically generated by a spatial light modulator, in order to acquire the spatial information of the target. As a consequence, the spatial resolution of each pattern should be equivalent to the pixel resolution $p\times q$ of the target image $x$. By leveraging the fact that natural images can be sparsely represented in an appropriate basis $\Psi$, i.e., $x=\Psi x'$, compressive imaging methods allow one to reconstruct the images with a few patterns whose number is only a fraction of the number of pixels $N$. Here we define the sampling rate as the ratio of the number of measurements to the number of pixels. This sampling rate is much smaller than that prescribed by Nyquist sampling. It typically takes about $M=O(K\cdot\log(N/K))<N$ random patterns, considering all but the largest $K(\leq N)$ elements of the sparse representation coefficients in some basis to be set to zero. Thus CS in principle provides a benefit in reducing acquisition time. And generally, $\Psi$ is an invertible (e.g. orthogonal) matrix or a redundant dictionary. The image $x$ can be reshaped into a column vector of size $N\times1$, where $N=p\times q$. In CS, the patterns are modulated by a digital micromirror device (DMD) consisting of millions of micromirrors, each of which is orientated either $12^\circ$ or $-12^\circ$ with respect to the normal of the DMD work plane, corresponding to a bright pixel 1 or a dark pixel 0. Each pattern $a_{ij}$ sequentially encoded on the DMD can be flattened into a row vector of size $1\times N$, thus $M$ such binary patterns constitute a known $M\times N$ measurement (or sensing) matrix $A$. This measurement matrix actually projects the object signal $x$ into a single-pixel (bucket) compressed signal $y=Ax+e$ of smaller size $M\times1$. Here $e$ is of the same size $M\times1$, denoting the stochastic noise. Thereby, the single-pixel total intensity measurement is mathematically equivalent to the inner product between each pattern and the object image, i.e., the interaction between the pattern sequence and the scene. Then the goal is to solve such an ill-posed linear problem through optimization algorithms by finding the sparsest representation $x'$ such that $y=A\Psi x'+e$. In order to ensure a good estimation of $x'$, the measurement matrix $A$ should satisfy the restricted isometry property (RIP) \cite{Candes2008}, which requires the sensing matrix with the property that column vectors taken from arbitrary subsets are approximately orthogonal and incoherent with $\Psi$. As we know, the Hadamard matrices can fulfil this property. Here we apply the total variation minimization \cite{CBLi2010}, whose objective function can be written as an augmented Lagrangian function:
\begin{equation}%\label{}
\min\limits_{x}\sum\limits_i||D_ix||_p+\frac{\mu}{2}||y-Ax||_2^2,
\end{equation}
where $||x||_p=(\sum_{i=1}^{N}|x_i|^p)^\frac{1}{p}$, $D_ix$ denotes the discrete gradient vector of $x$ at the $i$th position, $D$ is the gradient operator, and $\mu$ is a balance constant. Here a TVAL3 solver \cite{CBLi2010} is used to recover the image.

In this work, we make use of the Hadamard matrices to form our patterns. A Hadamard matrix is named after the French mathematician Jacques Salomon Hadamard. It is a symmetric square matrix with entries $\pm1$. Let $H$ be a Hadamard matrix of order $N$, then we have $HH^T=NI_N$ and $H^T=H$, where $I_N$ is the $N\times N$ identity matrix and $T$ represents the transpose operator. Dividing $H$ by $\sqrt{N}$ gives an orthogonal matrix whose transpose equals to its inverse. The Hadamard matrix of order $2\leq2^k\in N$ can be given by the following recursive formula
\begin{equation}%\label{}
H_{2^k}=\left[{\begin{array}{*{20}{c}}
H_{2^{k-1}}&H_{2^{k-1}}\\
H_{2^{k-1}}&-H_{2^{k-1}}
\end{array}}\right]=H_2\otimes H_{2^{k-1}},
\end{equation}
where $H_1=[1]$, $H_2=\left[{\begin{array}{*{20}{c}}
{1}&{1}\\
{1}&{-1}
\end{array}}\right]$, $\otimes$ stands for the Kronecker product. Such natural ordered Hadamard matrix is also called Walsh matrix. It is interesting to note that the elements in the first row and the first column are all ones.

\begin{figure}[H]%[htbp]
\centering\includegraphics[width=0.95\linewidth]{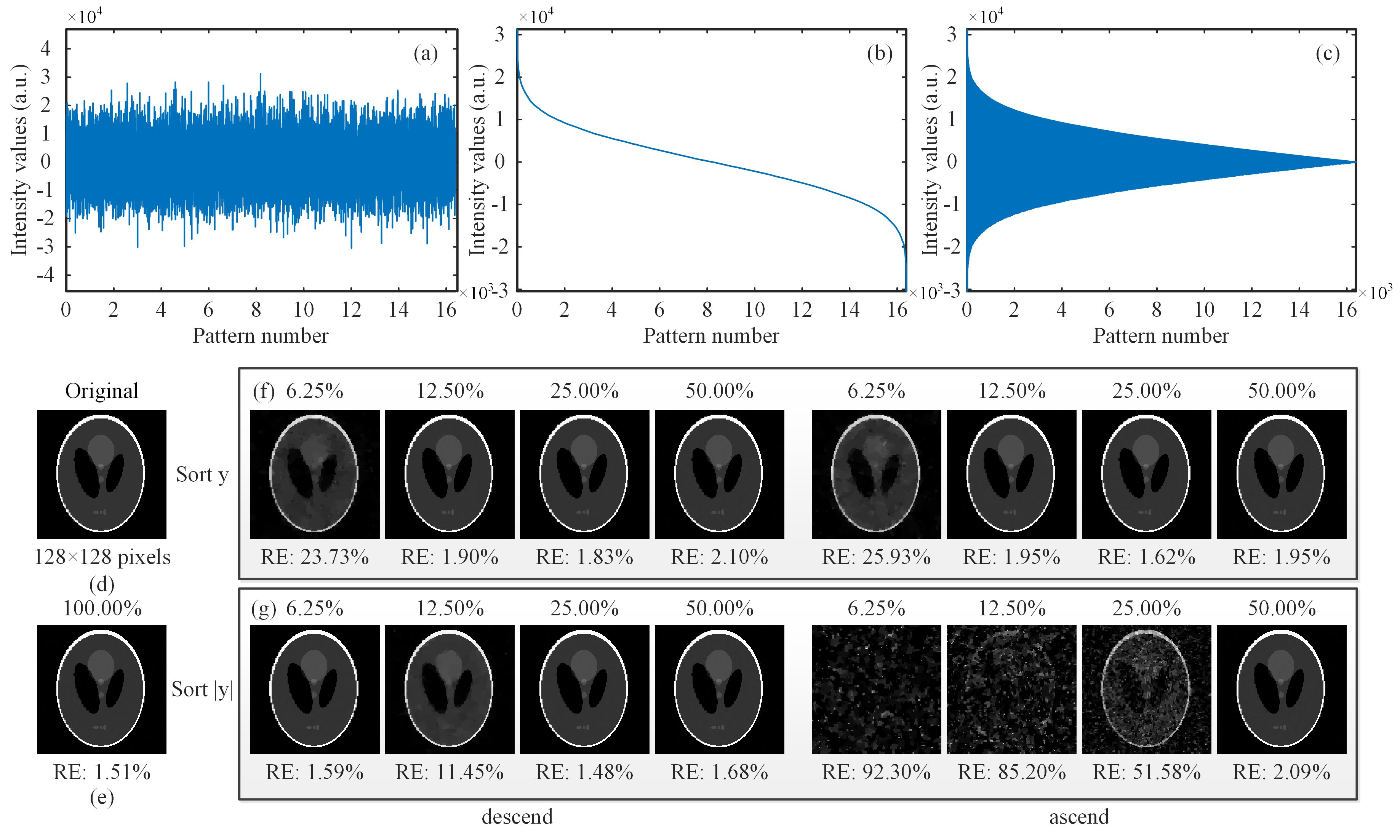}
\caption{Image reconstructions with different subsets of Hadamard patterns. (a) The original distribution of the bucket intensity signal $y$. (b) and (c) gives the intensity distribution when $y$ and $|y|$ is in a descending order, respectively. (d) is an original image of $128\times128$ pixels. (e) CS reconstruction using a complete set of the random patterns. (f) and (g) Comparison of CS reconstructions with the compression ratios of 6.25\%, 12.50\%, 25.00\% and 50.00\%, while the signal $y$ (or $|y|$) is in their descending order and ascending order, respectively.}
\label{fig:y}
\end{figure}

In CI, the patterns can be easily divided into a positive and a negative subset according to the bucket intensity fluctuation $\Delta y_i=y_i-\langle y\rangle$, expressed as
\begin{equation}%\label{}
A^+=\{A_i|\Delta y_i\geq0\}\ \textrm{and}\ A^-=\{A_i|\Delta y_i<0\},
\end{equation}
where $\langle\cdot\rangle$ signifies the ensemble average. A positive (or negative) correspondence image can be recovered by only averaging some fractions of the subset $A^+$ (or $A^-$). From this theory, it can be seen that the single-pixel (bucket) intensity is proportional to the contribution of the pattern to the image reconstruction. Now we plot the ordinary intensity signal $y=Ax$ in Fig.~\ref{fig:y}(a). Here the measurement matrix $A$ is randomly disrupted from $H$ (i.e., the rows and columns of $H$ are all scrambled to generate the $A$). The intensity values show a positive and negative distribution due to the fact that the entries of $A$ are either $+1$ or $-1$. Then we sort the signal $y$ and its absolute value form $|y|$ in a descending order, as displayed in Figs.~\ref{fig:y}(b)--\ref{fig:y}(c). By using the complete set of the patterns, the CS retrieved image is illustrated in Fig.~\ref{fig:y}(e). After that, we use some different sampling rates of 6.25\%, 12.50\%, 25.00\% and 50.00\% for CS image reconstruction, whilst the intensity values $y$ (or $|y|$) are in a descending order and a ascending order, respectively. The front part of Figs.~\ref{fig:y}(f)--\ref{fig:y}(g) show the retrieved images using the most significant fractions. From the results given in the latter part of Figs.~\ref{fig:y}(f)--\ref{fig:y}(g), we can see that, by using CS, the first few fractions of the subset of $A$ corresponding to $|y_i|-\langle|y|\rangle<0$, compared with the ones for $\Delta y_i<0$, present some poorer quality since they are unessential patterns to be more precisely. The results along with the relative errors (RE) demonstrate that, with the same sampling ratio, the significant fractions of the subset of $A$ with respect to $|y_i|-\langle|y|\rangle>0$ yield a better performance (see the front part of Figs.~\ref{fig:y}(f)--\ref{fig:y}(g)). In consequence, the most significant patterns should be modulated first. Although there exists many orders for the Hadamard matrix, such as sequency order, dyadic order, and so forth, none of them can make the most significant patterns appear first. Since it is hard to know \textit{a priori} which patterns can generate the most significant intensity values, one must perform a complete sampling and then pick up the crucial patterns needed according to the recorded signal.

The RD ordering \cite{MJSunSR2017} of the Hadamard basis provides an alternative GI approach. It numbers each row. The rows the Hadamard matrix are ordered such that the top half of $H_{2^{2z}}$ equals to the rows of $H_{2^{2z-1}}$ (in the two-dimensional pattern view, the former is an scaled version of the latter with a factor 2), just like a Russian dolls set, then the third quarter of $H_{2^{2z}}$ are ordered as the transpose of its second quarter, the rest is catalogued into the fourth quarter, and at last the patterns within each quarter are reordered again. By this means, the image can be reconstructed from a subset (e.g. 6\% sampling ratio) of significant patterns, with a quality comparable to the one of CS. And there is no need to disorder internal pixel layout in each pattern. But the RE of RD-based GI method presents a sawtooth descent as the sampling ratio increases, so that its performance curve is neither stable nor smooth. Since this method is based on the second order correlation, it is sensitive to the environmental noise. Additionally, the pixel resolution is also limited, generally $128\times 128$ pixels, as this ordering operation is too complex. Meanwhile, once the sampling ratio is fixed, the number of patterns required is determined by the total number of pixels of the reconstructed image. Therefore, it leads to a long acquisition time especially in imaging for large pixel resolution.

\noindent\textbf{Cake-cutting Hadamard basis sort.}
In this work, we propose a cake-cutting (CC) strategy to generate an optimized sort of the Hadamard basis. At first, each row of the Hadamard matrix $H$ is reshaped into a matrix of $p\times q=N$ pixels. Imagine each reshaped basis pattern as a cake, we can count how many pieces this cake is cut into. One piece of the cake can be defined as a connected region. In topology and mathematics, a connected region is a topological region that cannot be represented as the union of two or more disjoint nonempty subsets. This suggests each piece of the cake being either all $-1$ (in black) or $1$ (in white). Thus the piece number of the cake can be denoted by the number of connected regions of $-1$ plus those of $1$. Besides, for one pixel in one basis pattern, its adjacent pixels (up and down, left and right) with the same values can be all treated as some part of its connected region. According to the CI theory described above, only a small fraction of the complete patterns contributes to a larger intensity value $|y_i|$. Here we find that the less connected regions a pattern contains, the more probability of this pattern to be significant or to generate a higher measured value for a common object. Therefore, we order the complete Hadamard basis patterns according to their piece number and acquire a sort sequence $seq$ of size $N\times1$. After that, the $N$ reordered patterns can be flattened into $N$ row vectors, each of size $1\times N$, forming a $N\times N$ measurement matrix.

\begin{figure}[H]%[htbp]
\centering\includegraphics[width=0.95\linewidth]{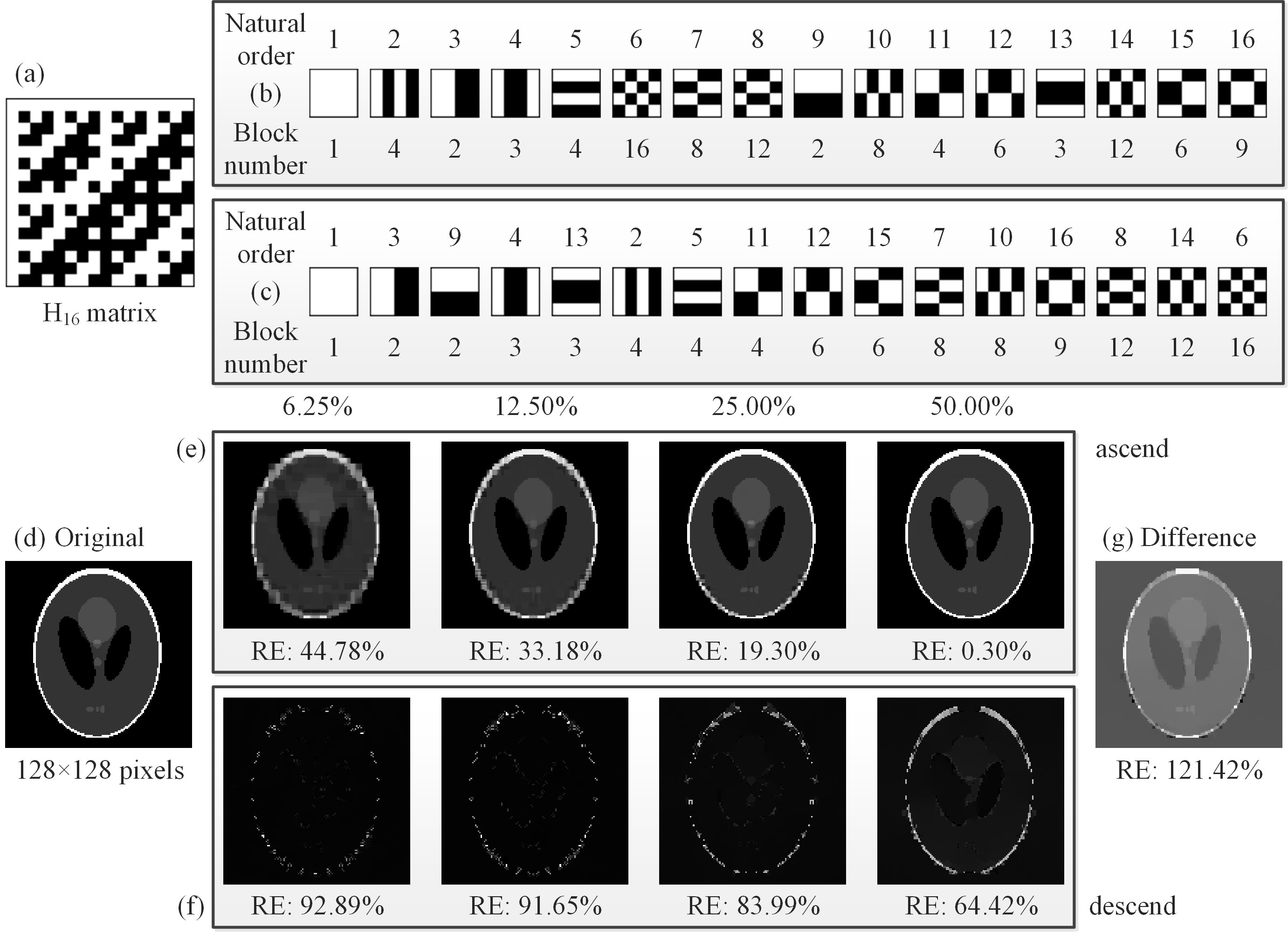}
\caption{An example for the ``cake-cutting" Hadamard basis. (a) A $16\times16$ Hadamard matrix $H_{16}$. (b) The basis patterns of $H_{16}$ in natural order. (c) The basis patterns of $H_{16}$ in the optimized ``cake-cutting" order. (d) An original head phantom image of $128\times128$ pixels. (e) and (f) Comparison of CS reconstructions with the first 6.25\%, 12.50\%, 25.00\% and 50.00\% of the fully sampled measurements while the piece numbers are in their ascending order and descending order, respectively. (g) The difference image of the fourth images of (e) and (f).}
\label{fig:Hadamard}
\end{figure}

Here, Fig.~\ref{fig:Hadamard} gives an example how our cake-cutting Hadamard basis sort works. By picking out each row of the $H_{16}$ matrix (Fig.~\ref{fig:Hadamard}(a)) and transforming each row into a $4\times4$ 2D pattern, a complete set of $16$ Hadamard basis patterns is then presented in Fig.~\ref{fig:Hadamard}(b), e.g., in natural order. Following our cake-cutting strategy, these Hadamard basis patterns are sorted by their piece numbers (see Fig.~\ref{fig:Hadamard}(c)). After that, we rebuild a Hadamard matrix $H_{16384}$ of order $N=128\times 128$, which is used to reconstruct a image of $128\times 128$ pixel resolution. By using our method, we compare the performance of recovered images from the first 6.25\%, 12.50\%, 25.00\% and 50.00\% of the complete measurements when the piece (block) number of each pattern are in ascending order and descending order, respectively, as shown in Figs.~\ref{fig:Hadamard}(e)--\ref{fig:Hadamard}(f). The results fit quite well with the theory. From the RE data, we can see that the performance is in proportion to the compression ratio for the ascending order but this feature is not suitable for the descending order. It is easy to think that the difference of the corresponding recovered images will produce a better quality, but it does not turn out that way due to the environmental noise, as shown in Fig.~\ref{fig:Hadamard}(g).

\begin{figure}[H]%[htbp]
\centering\includegraphics[width=0.95\linewidth]{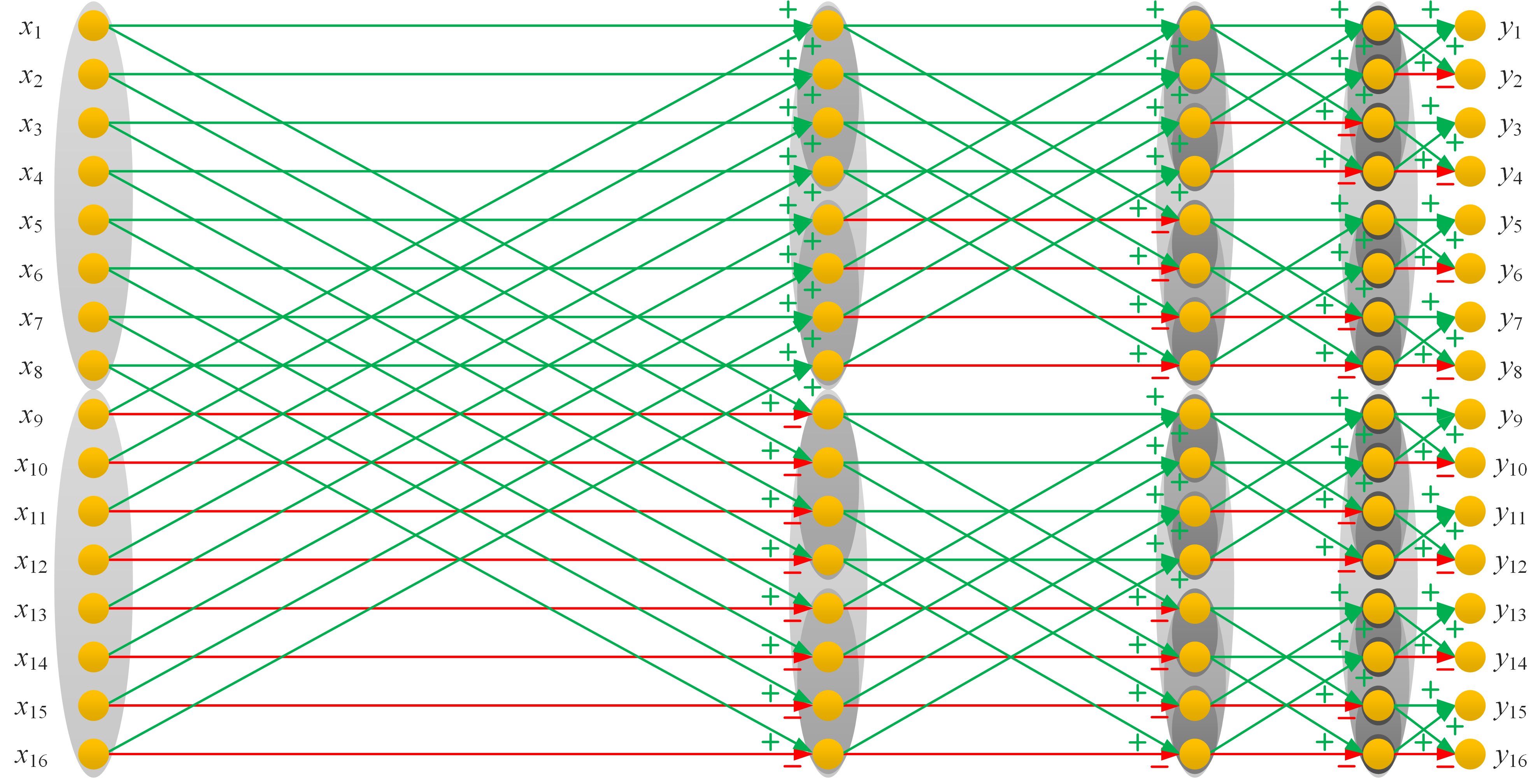}
\caption{Simplified mathematical model (graph) for the Hadamard matrix multiplication calculation, $H_{16}$ for example.}
\label{fig:Calculation}
\end{figure}

\noindent\textbf{Fast Hadamard computation.}
Now let us recall the structured characteristic of the Hadamard matrix. In computational mathematics, the Hadamard ordered fast Walsh-Hadamard transform (FWHT) is an efficient algorithm which can reduce the computational complexity of original $n$-order Walsh-Hadamard transform (WHT) from $O(n^2)$ to $O(n\log n)$, where $O$ is short for the order. This algorithm is a divide and conquer algorithm which recursively divides a WHT problem of size $n$ into two smaller WHT sub-problems of size $n/2$ \cite{Fino1976}. The idea of the FWHT applied to a column vector of length 16 is illustrated in Fig.~\ref{fig:Calculation}. Actually, the operation $H_Nx=y$ can also be explained by a graph with a set of vertices of edges. The weight of each edge is either $1$ or $-1$. Just like the neural networks or convolutional neural networks, the Hadamard matrix can be regarded as a propagation function or a network consisting of connections, in which each connection transfers the output of a neuron $i$ to the input of another neuron $j$, whilst $x=\{x_1,x_2,x_3,\ldots,x_N\}^T$ and $y=\{y_1,y_2,y_3,\ldots,y_N\}^T$ can be treated as the original input and the final output. There should be $\log_2N-1$ hidden layers, depending on the order $N$ of the Hadamard matrix. Additionally, there are $N$ transverse edges and $N$ intersection edges from the current layer to the next layer, and the number of the neurons in each layer is all $N$. Form the graph, it is interesting to find that all oblique intersection lines are in green, and all the lines that point to the right are half in green and half in red, where green stands for plus and red is minus.

For the Hadamard matrix $H$ of order $N$, the number of the computation layers of the fast computation for the formula $H_Nx=y$ should be $\log_2N$. First of all, we initialize an intermediate column vector $b$ such that $b=x$ and let $t=N/2$ be the original length of each group in the first layer (i.e., the input layer). For the $i$th layer expect for the output layer, i.e., $i=1,2,3,\ldots\log N$, the current $x$ can be divided into $2^i$ groups. In the $i$th layer, we will traverse $2^{i-1}$ times to compute every element in each group such that $temp=b(index)$, $b(index)=temp+b(index+t)$ and $b(index+t)=temp-b(index+t)$, where the range of $index$ is from $1+2(j-1)t$ to $(2j-1)t$, $j$ is from 1 to $2^{i-1}$, and $t$ denotes the length of each group in this layer. In the next $(i+1)$th layer, we update $t$ via $t=t/2$, repeat the above operations until the $(\log_2N+1)$th layer is reached. At last, $y=b$. If we want to pick out the $r_i$th row of the Hadamard matrix to form a modulated pattern, it just need to compute $H_Nx$ where the $r_i$th element of $x$ is set to one with the rest elements all zeros. The operation $H_N^{-1}x$ is equivalent to calculate $\frac{1}{\sqrt{N}}H_N^Tx=\frac{1}{\sqrt{N}}H_Nx$. Similarly, the $r_i$th element of $y$ can be also easily obtained after performing the above graph calculation. The sequency ordered (also known as Walsh ordered) FWHT, is generated by computing the natural (Hadamard) ordered FWHT, and then rearranging the outputs. Therefore, here we perform our cake-cutting strategy on the Walsh ordered operator, and choose the front $M$ elements of the output signal $y$ following the cake-cutting sort sequence. It should be noted that if the operator is chosen as other kind of FWHT, like the dyadic (Paley) ordered FWHT, the natural (Hadamard) ordered FWHT, the CC method should also be applied on the corresponding operator, otherwise the optimized order will be incorrect. Based on the above rules, the computation can be greatly simplified. Since there is no need to store the sensing matrix again, the memory consumption is also dramatically reduced.

\begin{figure}[H]%[htbp]
\centering\includegraphics[width=0.95\linewidth]{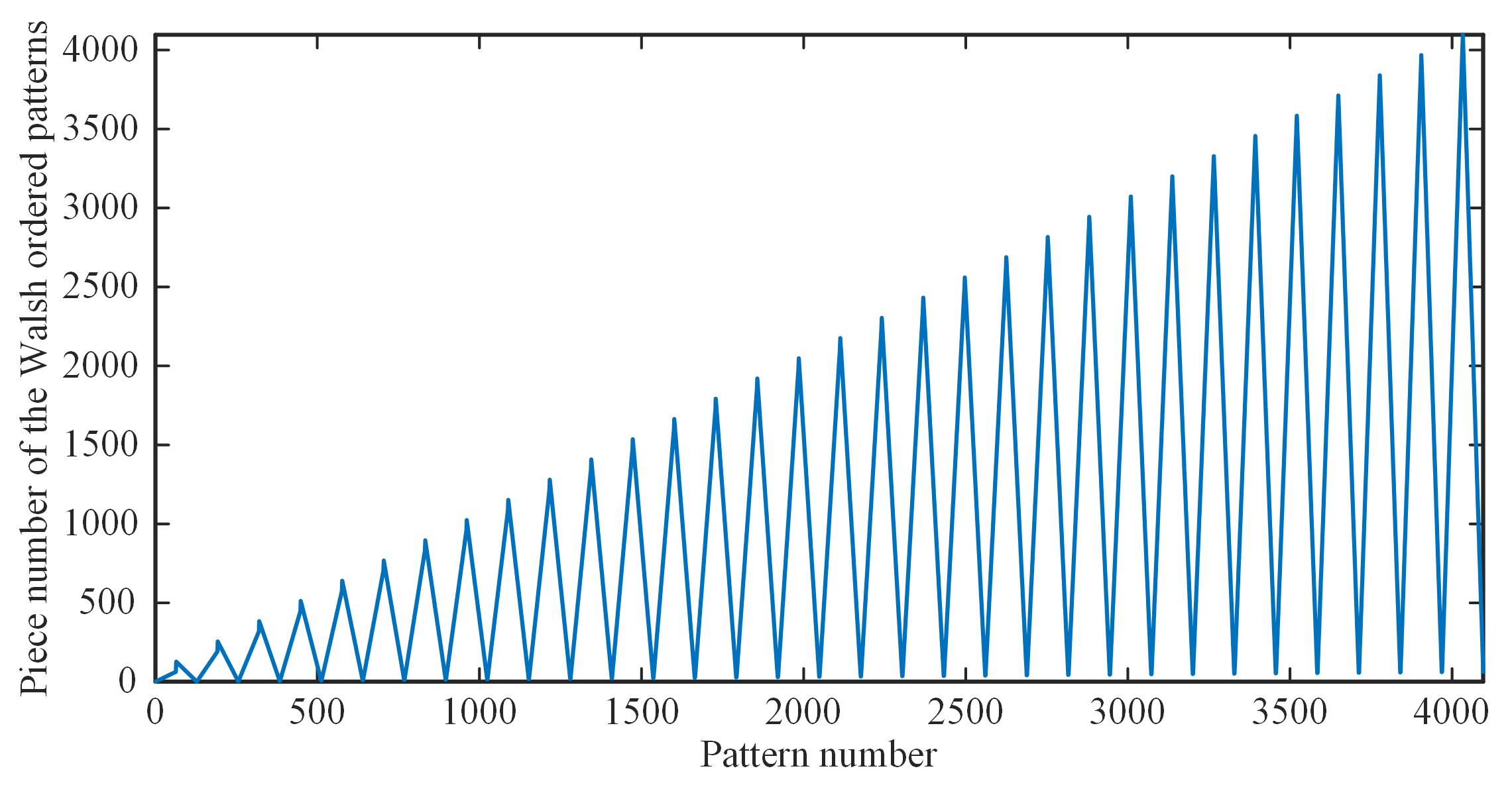}
\caption{The piece number of the Walsh ordered patterns as a function of the pattern number.}
\label{fig:rule}
\end{figure}

As for the order sequence generation time of our method, we present the comparison result in Table~\ref{tab:time}. Here we hypothesize that the image to be reconstructed is square, i.e., $p=q$. Since there is no need for the nested grouping like in RD method, our approach can greatly reduce the generation time of the sort sequence, especially for the large scale Hadamard matrix. But we still think that computing the number of the connected regions is time consuming (see the fourth row in Table~\ref{tab:time}), thus we plot the curve between the piece number of the Walsh ordered patterns and the pattern number, as shown in Fig.~\ref{fig:rule}. Fortunately, we find that there exists a remarkable regularity for $i=1,2,3,\ldots,q$, that is
\begin{equation}%\label{}
\left\{\begin{array}{l}
Seq\left[{(i-1)p+1:ip} \right]=i:-{(-1)^{mod(i,2)}}i:ip,{\textrm{\ for\ }}i{\textrm{\ is\ odd,}}\\
Seq\left[{(i-1)p+1:ip} \right]=ip:-{(-1)^{mod(i,2)}}i:i,{\textrm{\ for\ }}i{\textrm{\ is\ even.}}
\end{array}\right.
\end{equation}
By this means, we provide the order sequence generation time of $\textrm{CC}_{\textrm{rule}}$ for different $n$ in the fifth row of Table~\ref{tab:time}. From the data, we can see that this rule can further dramatically reduce the generation time of our CC method and makes our method more practical. Specifically, this regularity is only effective for the Walsh ordered patterns.

\begin{table}[ht]
\centering%\fontsize{8}{8}\selectfont
\caption{Comparison of the order sequence generation time (s) for RD and our cake-cutting method.}
\vspace{0.15cm}
\begin{tabular}{cccccccc}
\hline\hline
$n$&64&256&1024&4096&16384&65536\\
$p=q$&8&16&32&64&128&256\\
\hline
RD&0.0126&0.0957&2.2301&103.0852&5818.4470&too long\\
CC&0.0052&0.0232&0.1629&2.0058&33.8586&3220.034\\
$\textrm{CC}_{\textrm{rule}}$&0.000046&0.000051&0.000055&0.000111&0.000224&0.000562\\
\hline\hline
\end{tabular}\label{tab:time}
\end{table}

\begin{figure}[H]%[htbp]
\centering\includegraphics[width=0.95\linewidth]{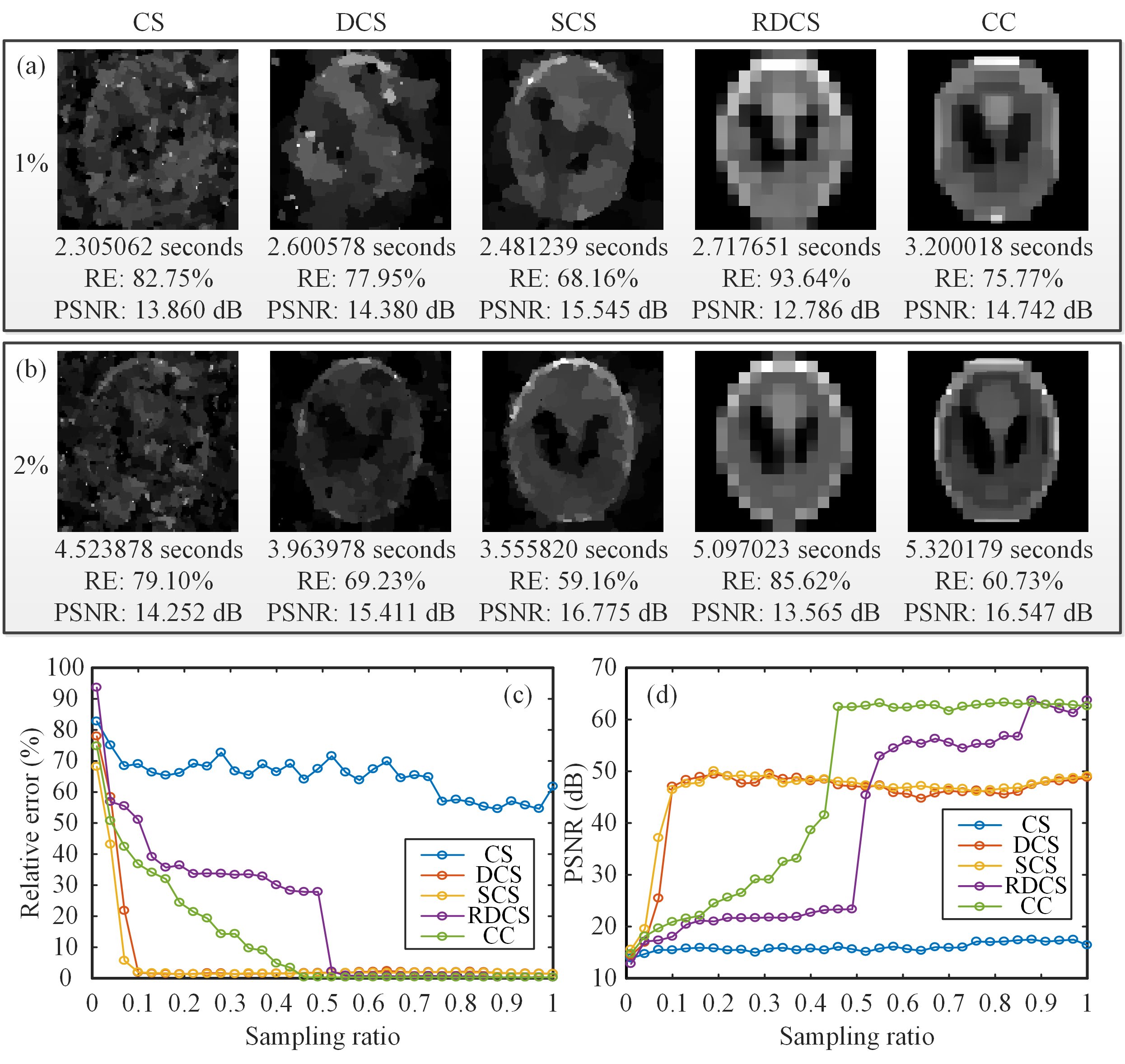}
\caption{Comparison of the simulation results. (a) and (b) show the reconstructions using compressed sensing (CS), differential compressed sensing (DCS), sorted compressed sensing (SCS), ``Russian Dolls" compressed sensing (RDCS), our ``cake-cutting" (CC) method, with a sampling rate of 1\% and 2\%, respectively. The corresponding evaluation parameters like the recovery time, relative error (RE) and peak signal-to-noise ratio (PSNR) are also given here. (c) and (d) give the comparison of above methods, in terms of RE along with PSNR as a function of the sampling ratio.}
\label{fig:Performance}
\end{figure}

\noindent\textbf{Numerical simulations.}
In order to test the performance of our method for image reconstruction, some numerical simulations are performed. Here we first create a head phantom image as the original image, which is normalized to a range of $0\sim255$. The results are acquired from 1\% and 2\% measurements by using five different approaches: CS, differential compressed sensing (DCS), sorted compressed sensing (SCS) (like in Fig.~\ref{fig:y}(g)), ``Russian Dolls" CS (RDCS), and our ``cake-cutting" method, as illustrated in Figs.~\ref{fig:Performance}(a)--\ref{fig:Performance}(b). It is worth mentioning that the ``Russian Dolls" method used here is applied to compressive imaging, rather than ghost imaging in the original scheme \cite{MJSunSR2017}, definitely generating a better image quality. It only takes a little more (negligible) time for our ``cake-cutting" method to iteratively compute the images, but yields a much better performance, compared with the other existing methods. Then we draw the RE and the peak signal-to-noise ratio (PSNR) of reconstructed images as a function of sampling ratio. From Figs.~\ref{fig:Performance}(c)--\ref{fig:Performance}(d), it is clearly seen that our CC method is much better than the CS and RD methods with an overwhelming superiority for any sampling ratio, and will exceed the DCS and SCS methods for the sampling ratios over 40\%. As mentioned above, SCS has a major drawback that it needs to fully sample the image, and then to pick up the most significant intensities. Our CC method makes up for this defect. It is important to note that, RE and PSNR, served as the performance metrics, all quantify the visibility via the calculation of pixel errors. These pixel-wise performance measures may fail to capture the correlation structure of the natural images and may cause evaluation misjudgments; e.g., an image which is supposed to have a better visibility may instead have a worse RE or PSNR value. Thereby from Figs.~\ref{fig:Performance}(a)--\ref{fig:Performance}(b), the CC method actually has a much better image quality when the sampling ratio is very low, but cannot be characterized very well with the data of Figs.~\ref{fig:Performance}(c)--\ref{fig:Performance}(d).

\begin{figure}[H]%[htbp]
\centering\includegraphics[width=0.95\linewidth]{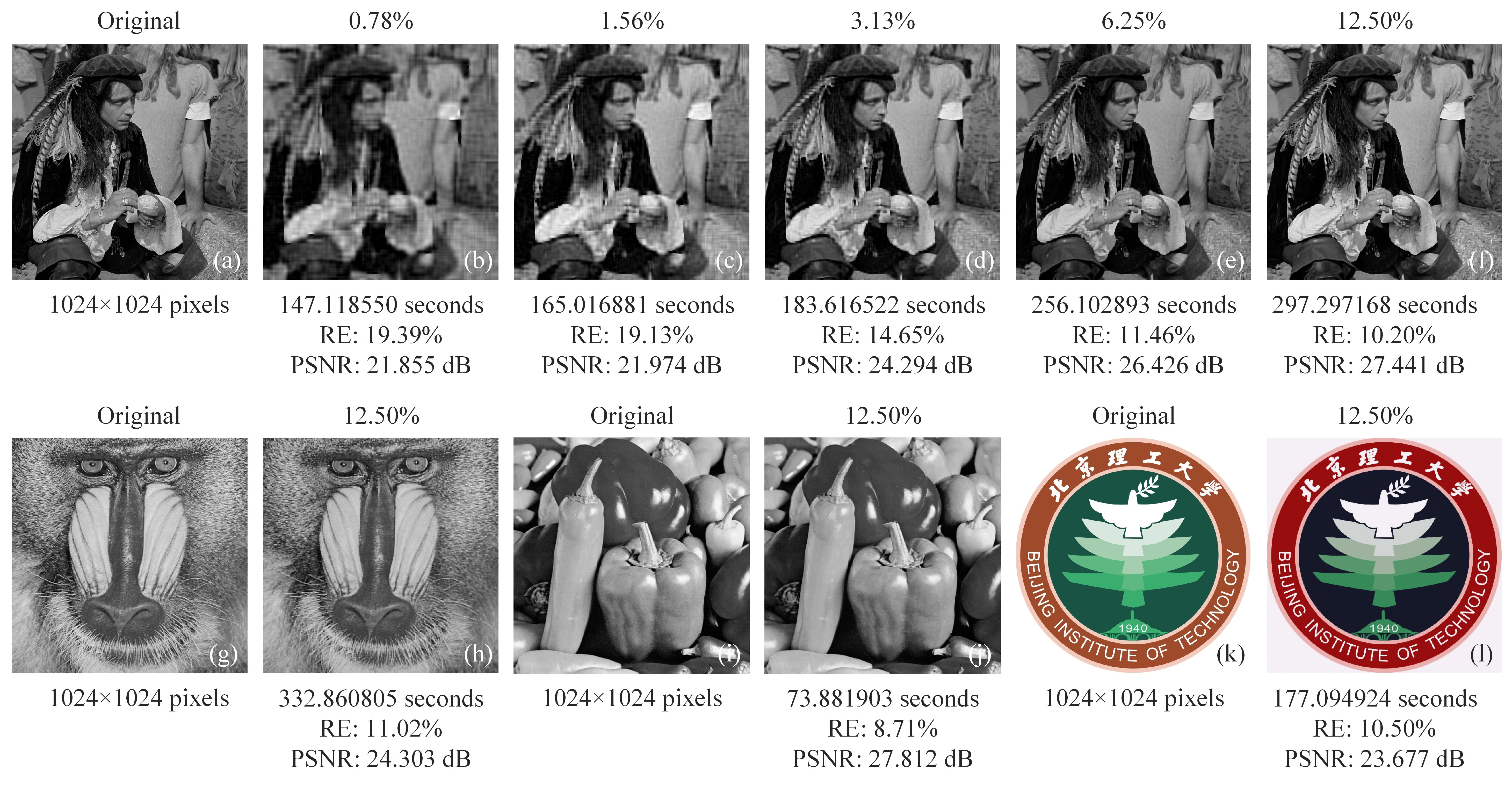}
\caption{Reconstructions of different large-pixel-size 2D images, covering gray-scale and color-scale, all of $1024\times1024$ pixels. (a) An original man image. (b)--(f) Retrieved images with a $2\times$ stepping increase of the sampling ratio. (g) An original mandrill image. (h) Recovered image of the mandrill with a sampling ratio of 12.50\%. (i) An original Lena image. (j) The reconstructed image of Lena using 12.50\% measurements. (k) The original color image of the school badge of Beijing Institute of Technology. (l) The reconstruction of the school badge with a compression ratio of 12.50\%. All original images are open access. We acknowledge Beijing Institute of Technology for the permission of using the school badge as an experimental object.}
\label{fig:Large}
\end{figure}

Next, another simulation is made to see the applicability of our method for the object images of large pixel scale. The original gray-scale objects are chosen from the open access standard test image gallery. Here the images of the man (Fig.~\ref{fig:Large}(a)), the baboon (Fig.~\ref{fig:Large}(g)) and Lena (Fig.~\ref{fig:Large}(i)) are used, all have the same resolution of $1024\times1024$ pixels. The reconstructions of the man image are performed at sampling ratio set from 0.78\% to 12.50\%, with a $2\times$ stepping increase, as shown in Figs.~\ref{fig:Large}(b)--\ref{fig:Large}(f). Since these results are retrieved with the same CC method, it is okey to use the RE and PSNR as the quality evaluation criterion. From the results we can see that the image quality and calculation time increases with the sampling ratio. Then the reconstructions (see Figs.~\ref{fig:Large}(h) and \ref{fig:Large}(j)) for different object images with 12.5\% sampling ratio aim to simulate the imaging of general scenes. For color-scale cases, we choose our school badge (Fig.~\ref{fig:Large}(k)) as the object, which will be split into the red, green and blue layers. By synthesizing the recovered images of the three wavelength components, the reconstruction of the color image can be obtained, as shown in Fig.~\ref{fig:Large}(l). The result shows that a multi-wavelength composite image can be reconstructed clearly with 255 tones with little color distortion. Above results are all performed with additive white gaussian noise (its mean is 1\% of the measured values mean, and its variance is 1).

\begin{figure}[htbp]%[H]
\centering\includegraphics[width=0.95\linewidth]{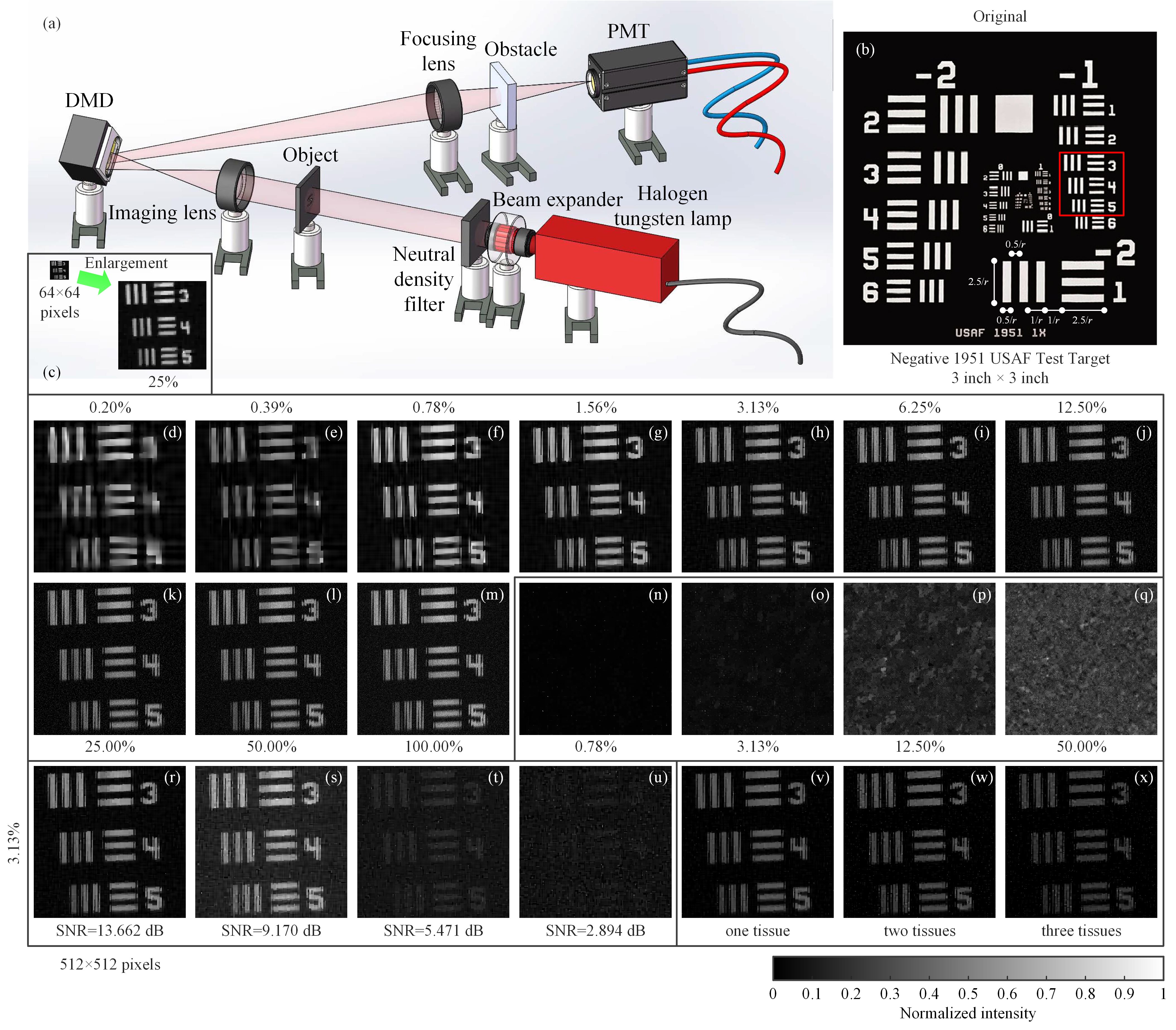}
\caption{Schematic of the experimental setup and experimental results. (a) The thermal light from a Halogen tungsten lamp illuminates the object through a beam expander and some neutral density filters, and then the light is projected onto a digital micromirror device (DMD). The reflected light is collected by a photomultiplier tube (PMT) through a focusing lens and some obstacles. (b) A negative 1951 USAF resolution test chart of 3~inch$\times$3~inch is treated as an object to be detected. (c) The recovered image of $64\times64$ pixel size with a sampling ratio of 25\%. (d)--(m) Reconstructed images of $512\times512$ pixels with a $2\times$ stepping increase of the sampling ratio, when the piece numbers are in their ascending order. (n)--(q) Retrieved images of $512\times512$ pixels with a $4\times$ stepping increase of the sampling ratio, when the piece numbers are in their descending order. (r)--(u) Reconstructions under different SNRs. It is noteworthy that the results of (c)--(u) are acquired without obstacles. (v)--(x) Results with increasing tissues as the obstacle. The third row results all use a sampling ratio of 3.13\%.}
\label{fig:Experiments}
\end{figure}

\noindent\textbf{Experimental setup and results.}
In our experimental setup, as shown in Fig.~\ref{fig:Experiments}(a), the object is illuminated by the collimated and attenuated thermal light beam emitted from a stabilized halogen tungsten lamp whose wavelength range covers from 360~nm to 2600~nm. Some 2 inch $\times$ 2 inch neutral density filters are used to attenuate the light to the ultra-weak light level. The transmission light from the object vertically incidents upon a DMD via a imaging lens. The reflected light from the DMD in $-24^\circ$ direction with respect to the normal incidence input beam is then sampled by a counter-type Hamamatsu H10682-210 photomultiplier tube (PMT). Since the PMT records the total intensity in the form of photon counts, it can be regarded as a single-photon single-pixel (bucket) detector. Our 0.7 inch DMD (ranged from 350~nm to 2700~nm) consists of $1024\times768$ pixels, each of size 13.68~$\mu$m$\times$13.68~$\mu$m. The states ``on" and ``off" of the micromirrors are determined by a preloaded sequence of binary patterns. The nominal maximal binary pattern switching rates of the commercially available DMDs reach $32550$~Hz (patterns/s) with an onboard storage for up to 45000 patterns. We have developed an improved DMD which enables us to load the pattern sequence onto the DMD in real time. We use the Hadamard basis in ``cake-cutting" sequence to generate our DMD modulated patterns.

The elements of the Hadamard matrix $A$ take values of 1 or $-1$, but the binary patterns encoded on the DMD consist of the values of 1 or 0, which can not ensure a good image quality with respect to the RIP. We found that by subtle shifting and stretching operations the matrix $A$ can be divided into two complementary matrices $\hat{A}=(A+1)/2$ and $\check{A}=1_{N\times N}-\hat{A}$, where 1 stands for an array consisting all ones. Thus the Hadamard matrix in optimal sort can be modulated by displaying each basis pattern of $\hat{A}$ immediately followed by its inverse (complementary) pattern (the one of $\check{A}$, i.e., the micromirror states ``on" and ``off" are reversed) on the DMD. Then the range of values of the differential patterns $A=\hat{A}-\check{A}$ becomes either 1 or $-1$, actually realizing ``positive-negative" intensity modulation and making the differential measurements with a mean $\sim0$. As a result, the SNR is also greatly improved \cite{YuOC2016,YuSR2014}.

For simplicity, we test our system by imaging a gray-scale object. Here we choose a negative 1951 USAF resolution test chart as the original object (see Fig.~\ref{fig:Experiments}(b)), whose black parts block the light and white parts transmit the light. The red square is the object image projected on the square central region of our DMD, covering $512\times512$ pixels (micromirrors). Note that the red suqare is in Group $-1$, thus the width of each line for Elements $3\sim5$ is 793.70~$\mu$m, 707.11~$\mu$m and 629.96~$\mu$m, respectively. The computation in detail is presented in the section of Methods.

In our experiments, we first modulate an optimized sequence of $64\times64$-pixel Hadamard basis patterns, where each pattern pixel comprises of $8\times8$ adjacent micromirrors. Without loss of generality, we make the DMD operate at 100~Hz. By utilizing our strategy, a coarse image of the object with a low-resolution of $64\times64$ pixels is retrieved from 1024 patterns, i.e., with a sampling rate of 25\%, as shown in Fig.~\ref{fig:Experiments}(c). By using a neutral density filter of transmissivity 0.001, the number of detected signal photons per image pixel is $\sim0.79$. Then Figs.~\ref{fig:Experiments}(d)--\ref{fig:Experiments}(m) illustrate $512\times512$ reconstructed images by using a series of sampling ratios with a coverage from 0.20\% to 100\%, when the piece numbers are in their ascending order. By contrast, Figs.~\ref{fig:Experiments}(n)--\ref{fig:Experiments}(q) show some examples of using the descending order of the Hadamard patterns, all with bad performances. Then we compare the results under different SNRs (see Figs.~\ref{fig:Experiments}(r)--\ref{fig:Experiments}(u)), by applying four different neutral density filters, whose transmissivity is 0.001, 0.0025, 0.005 and 0.01, respectively. Note that in above experiments there is no obstacle placed in the light path to block the detection light. Now, as shown in Fig.~\ref{fig:Experiments}(a), the PMT, whose photosensitive surface faces the scene, is covered with some pieces of lens cleaning tissues (some kind of organic fiber and can be treated as the obstacle here). The piece number of the tissues is increasing from 1 to 3 and the results are presented in Figs.~\ref{fig:Experiments}(v)--\ref{fig:Experiments}(x). Some part of the light reflected from the DMD passes through the tissues and is partly scattered. Thus the light collected by the PMT is a mixture of the direct light and the indirect light. In both SNR changing and partially obscuring cases, the numbers of measurements for the reconstructions of $512\times512$ images are all only 8192.

\section*{\textsf{Discussion and Conclusion}}
In our single-pixel imaging system, the noise cannot be neglected. There are many sources of noise, like the ambient illumination noise induced by the light source (with temperature drift), the dark noise of the counter-type PMT, the stray light reflected from the metal frame of the DMD, the specular and diffuse reflections from the metal surface under the intervals and flipping gaps of micromirrors (the latter is associated with the patterns as well), the stray light that bounces back and forth between the metal surfaces, and so forth. In addition, the SNR decreases as the transmissivity of neutral density filters is increased, as they are placed after the light source to attenuate the illumination. But the more the light is transmitted, the larger the stray light of the metal surfaces will be. Here the differential measurements we performed can also be used to average the variance of independently and identically distributed noise, thus significantly improving the SNR in the measurement process, as well as the imaging performance. Furthermore, the image quality can be also improved if the other kinds of stochastic noise is well suppressed.

The programmable DMDs operating in binary mode with pattern displaying rate of 32.55~kHz are very common. Since the pulse-pair resolution of the used PMT is 20~ns, thus its working frequency is not the limitation of the system. As a consequence, the performance of the system is mainly restricted by the modulation rate of the DMD. Even so, tens of thousands of patterns per second could enable 1.97~ms (in total) measurement at $64\times64$-pixel resolution and 377.51~ms (in total) sampling at $1024\times768$-pixel resolution, all with a sampling ratio of 0.78\% (incorporating two adjacent complementary patterns for differential measurements). Therefore, for relatively low-resolution applications, it allows video rate image acquisition. By making full use of the structured characteristic of Hadamard matrix, we have demonstrated that the proposed method is capable of reconstructing large pixel-resolution images with high performance using few computational overhead and memory consumption. Thus the realization of the system hardware in the future will bring the single-pixel imaging closer to practical applications, for instance, mobile phones, night vision goggles or satellites.

The proposed technique employs ``cake-cutting" strategy for the Hadamard basis ordering. We think that the optimized sort sequence of the patterns might be deterministically described in mathematics or could be generated by other new methods in the near future. In this case, the generation time of the sort sequence will be further shortened. It is worth noting that the orthogonality of Hadamard matrix is the key (but not the only) factor of fast computation and its combination with CS allows a perfect reconstruction from super sub-Nyquist measurements even in the presence of noise. The reconstructions of other orthogonal matrices or deterministic matrices will be our future work.

For full-colour imaging, one can use three spectral filters to restore the red, green, and blue sub-images, and then synthesize the three sub-images to a color image. Moreover, the operational spectrum of the DMD ranges from 350~nm to 2700~nm, allowing the proposed system to be extended to the non-visible region of the spectrum where the array detectors are not well developed, such as in the infrared or ultraviolet wavelengths. In these situations, it only needs to change the lens and the single-pixel detector to fit the corresponding wavelength.

In a nutshell, we propose a single-pixel compressive imaging method based on ``cake-cutting" Hadamard basis ordering, which is capable of precisely reconstructing images of large resolution up to $1024\times1024$ pixels from super sub-Nyquist measurements. The sampling ratio can be shortened to even 0.2\%, thus significantly reducing the acquisition time. According to the significant contribution of the deterministic Hadamard basis to the image reconstruction, a optimized sequence of patterns is obtained by directly making the internal piece numbers of basis patterns in their ascending order. By making full advantage of the structured characteristic of Hadamard matrix, the predetermined patterns can be loaded onto the DMD in real time, without the need to be all stored on the DMD. Additionally, in terms of computational efficiency, it also offers fast computation along with a small computational memory requirement (in computer), due to the simplified mathematical calculation model for the Hadamard matrix multiplication and the orthogonality of the Hadamard matrix. We have demonstrated this method with a single-photon single-pixel camera based on differential modulation of the DMD. The experimental results proved that our technique enables a good image reconstruction from indirect measurements through a partially obscuring scene in the presence of noise or under ultra-weak illumination. The technique can be easily extended to single-pixel imaging in other non-visible wavebands and offers an avenue to overcome the limitations existing in the recently introduced single-pixel imaging schemes.

\section*{\textsf{Methods}}
\textbf{Target object.} The 1951 USAF resolution test chart consists of a series of stripes decreasing in size, while the standard target element is composed of two sets of lines, each set is made up of three lines separated by spaces of equal width. Suppose $r$ to be the number of lines per millimeter, the parallel lines are $2.5/r$ millimeters long and $0.5/r$ millimeters wide with space $0.5/r$ millimeters wide between the parallel lines. The space between the vertical and horizontal lines is $1/r$ millimeters wide. The elements within a group are numbered from 1 to 6, which are progressively smaller. The group number covers from $-2$ to 7. The length of any target element line can be expressed as $2.5/2^{\textrm{Group}+(\textrm{Element}-1)/6}$mm, while the width equals to the length divided by 5, also is equivalent to $0.5/$Resolution (line pair$/$mm). Thus it is easy to compute the width of each line in the red square (all in Group $-1$), i.e., 793.70~$\mu$m for Element 3, 707.11~$\mu$m for Element 4, and 629.96~$\mu$m for Element 5.

\noindent\textbf{Data processing.} All the data was analyzed and processed with MATLAB R2018b (The MathWorks, Inc.). The ordering and reconstructions were performed on a standard desktop computer with an Inter Core i7-6700 CPU @ 3.40~GHz and a memory of 16~GB. If a supercomputer with parallel processing is used or the system hardware is realized, the computation time will be much shorter.

\noindent\textbf{Image analysis.} To obtain a quantitative measure of the image quality, the relative error (RE) is defined here as a figure of merit:
\begin{equation}
RE=\frac{||\tilde U-U_o||_F}{||U_o||_F}\times100\%,
\label{eq:RE}
\end{equation}
where $\tilde U$ denotes the reconstructed image and $U_o$ stands for the original image, all of $p\times q$ pixels. Here, $||X||_F$ is called Frobenius norm which can be defined as
\begin{equation}
||X||_F=\sqrt{\sum_{i=1}^p\sum_{j=1}^q|X_{ij}|^2}=\sqrt{\textrm{trace}(X^\ast X)}=\sqrt{\sum_{i=1}^{min\{p,q\}}\sigma_i^2},
\label{eq:Fnorm}
\end{equation}
where $\ast$ denotes the conjugate transpose operator, and $\sigma_i$ is the singular value of $X$.

Additionally, here we introduce another unitless performance measure, the peak signal-to-noise ratio (PSNR), which is defined as
\begin{equation}
\textrm{PSNR}=10\log(255^2/\textrm{MSE}),
\label{eq:PSNR}
\end{equation}
where $\textrm{MSE}=\frac{1}{pq}\sum\nolimits_{i,j=1}^{p,q}[U_o(i,j)-\tilde U(i,j)]^2$. The MSE describes the squared distance between the recovered image and the original image. Naturally, the larger the PSNR value, the better the quality of the image recovered.

In order to allow a fair comparison of the image quality, all the recovered images of Figs.~\ref{fig:Hadamard}, \ref{fig:Performance}--\ref{fig:Large} are normalized to a range of $0\sim255$. Since the optical experiments generally have no original image as a reference, the results of Fig.~\ref{fig:Experiments} are directly normalized to a range of $0\sim1$.

\section*{\textsf{Acknowledgments}}
This work is supported by the National Natural Science Foundation of China (61801022), the Beijing Natural Science Foundation (4184098), the National Key Research and Development Program of China (2016YFE0131500), the International Science and Technology Cooperation Special Project of Beijing Institute of Technology (GZ2018185101), and the Beijing Excellent Talents Cultivation Project - Youth Backbone Individual Project. The author warmly appreciate Xiao-Peng Jin for preparing Fig.~\ref{fig:Experiments}.

\section*{\textsf{Author Contributions}}
W.-K.Y. conceived the idea, designed and performed the simulations and experiments, collected and analyzed the data, and wrote this manuscript.

\section*{\textsf{Additional information}}
\textbf{Supplementary Information} accompanies this paper at \url{http://www.XX.com/XX}

\vspace{3mm}\noindent\textbf{Competing financial interests} The authors declare no competing financial interests.

\vspace{3mm}\noindent\textbf{Reprints and permission} information is available online at \url{http://XXX.XXXXXX.com/XXXX/}

\vspace{3mm}\noindent\textbf{How to cite this article} Yu, W.-K. Super sub-Nyquist single-pixel imaging by means of cake-cutting Hadamard basis sort. Xxx. Xxxxxx. X:XXXX doi: XX.XXXX$/$XXXXXXXXXX (20XX).
\end{document}